\PassOptionsToPackage{unicode}{hyperref}
\PassOptionsToPackage{hyphens}{url}
\documentclass[
]{article}
\usepackage[a4paper, total={6in, 8in}]{geometry}
\usepackage{multicol}
\usepackage{authblk}
\usepackage{amsmath,amssymb}
\usepackage{iftex}
\ifPDFTeX
  \usepackage[T1]{fontenc}
  \usepackage[utf8]{inputenc}
  \usepackage{textcomp} 
\else 
  \usepackage{unicode-math} 
  \defaultfontfeatures{Scale=MatchLowercase}
  \defaultfontfeatures[\rmfamily]{Ligatures=TeX,Scale=1}
\fi
\usepackage{lmodern}
\ifPDFTeX\else
\fi
\IfFileExists{upquote.sty}{\usepackage{upquote}}{}
\IfFileExists{microtype.sty}{
  \usepackage[]{microtype}
  \UseMicrotypeSet[protrusion]{basicmath} 
}{}
\makeatletter
\@ifundefined{KOMAClassName}{
  \IfFileExists{parskip.sty}{%
    \usepackage{parskip}
  }{
    \setlength{\parindent}{0pt}
    \setlength{\parskip}{6pt plus 2pt minus 1pt}}
}{
  \KOMAoptions{parskip=half}}
\makeatother
\usepackage{xcolor}
\usepackage{longtable,booktabs,array}
\usepackage{calc} 
\usepackage{etoolbox}
\makeatletter
\patchcmd\longtable{\par}{\if@noskipsec\mbox{}\fi\par}{}{}
\makeatother
\IfFileExists{footnotehyper.sty}{\usepackage{footnotehyper}}{\usepackage{footnote}}
\makesavenoteenv{longtable}
\usepackage{graphicx}
\makeatletter
\def\maxwidth{\ifdim\Gin@nat@width>\linewidth\linewidth\else\Gin@nat@width\fi}
\def\maxheight{\ifdim\Gin@nat@height>\textheight\textheight\else\Gin@nat@height\fi}
\makeatother
\setkeys{Gin}{width=\maxwidth,height=\maxheight,keepaspectratio}
\makeatletter
\def\fps@figure{htbp}
\makeatother
\usepackage{soul}
\ifLuaTeX
  \usepackage{selnolig}  
\fi
\IfFileExists{bookmark.sty}{\usepackage{bookmark}}{\usepackage{hyperref}}
\IfFileExists{xurl.sty}{\usepackage{xurl}}{} 

\providecommand{\keywords}[1]
{
  \small	
  \textbf{\textit{Keywords---}} #1
}

\title{Unmaking AI Imagemaking: A Methodological Toolkit for Critical Investigation}

\author[1]{Luke Munn}
\author[2]{Liam Magee}
\author[3]{Vanicka Arora}
\affil[1]{University of Queensland, Australia \authorcr l.munn@uq.edu.au}
\affil[2]{Western Sydney University, Australia \authorcr l.magee@westernsydney.edu.au}
\affil[3]{University of Stirling, United Kingdom \authorcr vanicka.arora@stir.ac.uk}

\date{July 2023}

\begin{document}

\maketitle

\begin{abstract}

  AI image models are rapidly evolving, disrupting aesthetic production in
  many industries. However, understanding of their underlying archives,
  their logic of image reproduction, and their persistent biases remains
  limited. What kind of methods and approaches could open up these black
  boxes? In this paper, we provide three methodological approaches for
  investigating AI image models and apply them to Stable Diffusion as a
  case study. \emph{Unmaking the ecosystem} analyzes the values,
  structures, and incentives surrounding the model's production.
  \emph{Unmaking the data} analyzes the images and text the model draws
  upon, with their attendant particularities and biases. \emph{Unmaking
  the output} analyzes the model's generative results, revealing its
  logics through prompting, reflection, and iteration. Each mode of
  inquiry highlights particular ways in which the image model captures,
  ``understands,'' and recreates the world. This accessible framework
  supports the work of critically investigating generative AI image models
  and paves the way for more socially and politically attuned analyses of
  their impacts in the world.
  
\end{abstract}


\pagebreak

\begin{multicols*}{2}

\hypertarget{introduction-approaching-ai-image-models}{%
\subsection{Introduction: Approaching AI Image
Models}\label{introduction-approaching-ai-image-models}}

AI image models are rapidly evolving, disrupting aesthetic production in
many industries. AI image models are already being used in design as a
rapid prototyping tool (Kulkarni et al. 2023), in the medical space as a
way to synthesize and identify tumors (Park et al. 2022), and in
scientific research to generate remote sensing datasets (Abduljawad and
Alsalmani 2022). In this paper we focus on a broad array of emerging
text-to-image deep learning models that generate images based on user
``prompts.'' If these image models have enormous power and promise, they
also pose new problems and challenges. Research has already identified
how these image models internalize toxic stereotypes (Birnhane 2021) and
reproduce forms of gendered and ethnic bias (Luccioni 2023), to name
just two issues.

As image models are rolled out in these high-stakes areas, it becomes
increasingly important to develop a critical understanding of their
operations, limitations, and potential impacts for society. Yet image
models and AI more broadly are often pervaded by an array of myths and
misconceptions (Emmert-Streib et al. 2020). The public's grasp of their
underlying archives, their logic of image reproduction, and their
persistent biases remains limited. This lack of understanding can be
partially attributed to the novelty of generative models. DALL-E was
only released in 2021, followed by Stable Diffusion, MidJourney,
Google's Imagen, and Parti in 2022. Critical AI research, while in some
senses prolific, has struggled to keep up with the intense pace of
technical development, with new models released in some cases on a
weekly basis.

Lack of understanding is also a natural byproduct of the inherent
opacity around models. Models can often be closed-source or proprietary:
access to their underlying data is limited, their precise parameters are
unknown, and the way in which human decisions and reinforcement learning
shape the performance of the model remain unclear (Facchini and Termine
2022). Even in the case of open source models like Stable Diffusion, as
we discuss in this paper, much of the ``pipeline'' from capital
investments and data and hardware accumulation to model training,
refinement and deployment is held in confidence by ``steering''
corporations like StabilityAI or, when made public, technically dense
and opaque.

What kind of methods and approaches could open up these black boxes? In
this paper, we propose a methodological ``toolkit'' for investigating
generative AI image models. This toolkit deliberately draws on a radical
mix of methods from different research disciplines to develop a more
holistic portrait of these models. In doing so, it engages in the
``methodological and interpretative pollution'' that Amanda Wasielewski
(2023, 138) asserts is necessary to engage with the diversity of
practices of image making that proliferate this space. These models are
products that emerge from a particular commercial \emph{ecosystem}
composed of investors, business objectives, organizational incentives,
and so on (Luitse and Denkena 2021). But these image models should also
be understood as \emph{datasets}, with embedded representations of
relations between textual tokens and pixel clusters which deeply shapes
their biases and functionality (Birhane et al. 2021). And finally, these
models produce \emph{output} which can be investigated through prompting
and iterating via the interface, revealing certain tendencies and logics
(Gal et al. 2022).

We want to remain attuned, then, to business structures but also data
structures. We want to grasp technical logics while remaining alive to
their racial, cultural, and political impacts. This multifaceted blend
echoes other research that combines analyses of a platform's political
economy with technical analyses of its affordances and outputs (Nieborg
and Helmond 2019) or understands these objects as an interplay of
technologies, tasks, structures, and actors (Ghaffari et al. 2019). In
providing this blend, we aim to support a socioculturally nuanced
analysis, elevating our technical literacy and critical awareness of AI
image models and their increasingly significant impacts in the world.

To illustrate our trio of methods, we use Stable Diffusion as a case
study. Since its release in 2022, Stable Diffusion has quickly risen to
prominence, creating immense amounts of public discourse and media
attention (Clarke 2022; Myer 2022; Hill 2023). We chose Stable Diffusion
because of this massive attention, but also because of its influence
over AI image generation, its open-source status, its relatively
accessible developer community, and its largely transparent updates (as
opposed to proprietary, closed-source offerings like DALL-E or
MidJourney). Stable Diffusion, then, is both important and
accessible---and this makes it an ideal object of research to test
methods against.

We stress however, that Stable Diffusion is just one possible model
amongst many, highlighting the need for a flexible and adaptable
methodological toolkit. For this reason, the three ``methods'' outlined
here are better understood as modes of inquiry rather than step-by-step
recipes. For example, AI image models differ significantly in how much
they disclose about their training data and its provenance, as well as
the degree to which this can be investigated. Certainly then,
researchers will need to adapt their activities to a particular model or
object-of-inquiry. However, rather than starting from scratch, the
questions, concepts, and references used in our case study provide a
blueprint for how these inquiries might be carried out.

In the first section, we situate our intervention in relation to
research on AI image-making and critical AI methods more broadly. In the
following three sections, we apply these methods to Stable Diffusion as
a case-study. \emph{Unmaking the ecosystem} analyzes the values,
structures, and incentives surrounding the model's production.
\emph{Unmaking the data} analyzes the images-text pairings the model
draws upon, with their attendant particularities and biases.
\emph{Unmaking the output} analyzes the model's generative results,
revealing some of its logics through prompting, reflection, and
iteration. We conclude by suggesting how these methods might be taken up
to advance nascent research agendas on AI image models.

\hypertarget{literature-review-situating-ai-imagemaking}{%
\subsection{Literature Review: Situating AI
Imagemaking}\label{literature-review-situating-ai-imagemaking}}

How might we understand the role of generative image models in reshaping
image production and visual culture? Here we draw together two strands
of scholarship. The first strand emerges from media studies, software
studies, and critical AI studies. It tends to highlight the novelty of
these technical regimes and investigate the algorithmic logics of these
systems to provide insights about their operations. The second strand
emerges from art history, art theory, and cultural theory, and focuses
on the production and curation of the image archive, its cultural,
ideological, and political impacts, and its shaping of the world and our
place within it. By splicing together these two strands, we approach the
generative model as both novel and historical, as both a technical
system and a sociocultural force.

For MacKenzie and Munster (2019), AI models assemble a certain
repository of images, quantified, labeled, and made platform-ready, into
the image ``ensemble.'' This ensemble, its embedding in countless model
variations, and its deployment in many different contexts creates a
visual regime that is too extensive, too overdetermined, to understand
through human perception alone. These conditions create the basis for a
new model of perception, a reformulation of visuality itself that they
term ``platform seeing'' (MacKenzie and Munster 2019).

This novelty can lead to unexpected results and disorienting
experiences. Offert and Bell (2021), for instance, investigate the
disconnect between the visuality of a machine vision system and more
conventional or historical modes of image production and perception.
They describe this discrepancy as a form of perceptual bias (Offert and
Bell 2021), a gap between our perceptual expectations and the perceptual
topology of the model. In this sense, this article takes up the
challenge of Azar et al. (2021), investigating how AI image models
unsettle the relation between what we see and what we know.

The artist Harun Farocki was prescient in demonstrating the novel logics
these visual regimes introduce. These ``operational images'' (Paglen
2014) work on two fronts. First, the image becomes a machinic output,
something assembled by a technical system, the end-result of a set of
algorithmic operations. The ``hand of the artist'' or photographer,
already pushed to the margins by successive stages of mechanical
reproduction, is almost entirely removed. Instead of an artisanal
process, we witness something wholly automated, a depiction arising out
of a set of computational process and combinatorial rules. Second, the
image becomes a kind of machinic input. As Paglen (2014) stresses, these
depictions are designed by machines and for machines rather than the
``meat eye'' of the human. They decenter the traditional human viewer,
with her tastes and aesthetic preferences, and instead privilege an
alternate set of priorities. Is the image (and any labeling)
mechanically legible? Can it be collected with millions of similar
images and cross-indexed to reveal correlations? Can it form a stable
ground or aesthetic foundations for subsequent image-production chains?
Here the image becomes a kind of engine, an informational space or data
package that only serves to reveal further insights, fine-tune
additional parameters, or generate more images.

However, if this first media studies strand excels in investigating the
technical logics of systems and their reshaping of visuality in society
broadly speaking, it can overlook the raw power of the image and its
influence on society, history, and culture more specifically. By their
very structure, notes Sekula (2003, 447), collections and archives of
images maintain a ``hidden connection between knowledge and power.'' The
archive, by definition, includes some images and excludes others; it
adopts a certain gaze and assigns value to certain people, places, and
things. And yet this archive is also a ``territory of images,'' (Sekula
2003, 444), a clearing house of depictions which is owned and operated
on. Depictions can be lifted out of this repository, fused with other
images, and adapted to tell a particular story, to document history in a
particular way, or to uphold a set of hegemonic values. As Derrida
(1996, 10) observed, the archive's powers of identification and
classification are also joined by the power of consignation, or the
gathering together of signs. These signs can be cut, spliced, and
merged, in countless permutations, to yield new sign-chains that say new
things and communicate new messages.

The two strands laid out here have begun to converge in recent
scholarship. Parisi (2021) asks how machine vision might move beyond the
ocular-centric Western gaze and the reproduction of racial capital.
Gil-Fournier and Parikka (2021) draw on historical satellite archives to
question the apparent stability of AI models and their ``ground
truths.'' Offert and Phan (2022) suggest that AI image models use latent
space to perpetuate established modes of racial politics and reproduce
whiteness as a dominant visual feature. And Wasielewski (2023) discusses
how artistic style is reshaped by deep learning and the new semantic
categories of large image-based datasets.

The generative image model provides a new way to mass-produce the image,
to synthesize infinite variations on an industrial scale by tweaking a
prompt. Such a development resonates with Benjamin's foundational essay
on the work of art in the age of mechanical reproduction. Benjamin
(1935) explored these new conditions at work in image cultures, showing
how the copied-and-circulated image became disconnected from its author,
its original context, and its attendant aura. However, rather than a
conservative lament, Benjamin's essay suggested this disconnect was
liberating, freeing up aesthetics to be consumed and appreciated in
novel ways. But as he and others (Crary 1992) have argued, each wave of
technical innovation and novel productive forces needs to be accompanied
by a corresponding societal adjustment: how to read these images and use
them to understand humanity's role in a technologically-mediated world.
Just like photography and film, then, we suggest that AI image models
``train human beings in the apperceptions and reactions needed to deal
with a vast apparatus whose role in their lives is expanding almost
daily'' (Benjamin 1935, 26).

While such a phenomenon undoubtedly raises many questions, immediate
among them is the question of method. Conventional means of examining
archives, of close-reading images, or of deconstructing visual symbolism
are only of partial help here. As Lev Manovich (2017a; 2017b) stressed
early on, algorithmic rules, automated processes, big data, and
networked media were going to fundamentally alter visual production,
leading to the emergence of new image cultures. These new sociotechnical
forms would require new methods of exploration, new approaches to get
``under the hood'' and grapple with their unfamiliar modes of
classification, retrieval, and synthesis. This article offers some tools
for dealing with the sociocultural apparatus of AI image models that
increasingly permeate everyday life.

\hypertarget{unmaking-the-ecosystem}{%
\subsection{\texorpdfstring{Unmaking the Ecosystem
}{Unmaking the Ecosystem }}\label{unmaking-the-ecosystem}}

To begin, we focus on unmaking the ecosystem---a metaphor we use to
describe the media ecologies and technical systems that surround the
model and bring it into existence. Encompassing financial,
organizational, technological, and cultural dimensions of the conditions
of model production, ecosystems do not have ready-made boundaries, and
so ``unmaking'' cannot provide a comprehensive account of those
conditions. However, even partially grasping this dense web of corporate
structures, capital investments, computer hardware, development teams,
and online communities can provide insights into the model.

This ecosystem strongly shapes the model, imposing a particular set of
aims, tendencies, and imperatives that are seen as desirable by the
model's various stakeholders (Woodhouse and Patton 2004; Vertesi et al.
2016). These differ between models too, as stakeholders make decisions
about licensing terms (open source or otherwise), platforms, access
rights, data sets, training algorithms, moderation policies, and so on.
As with the other methods, we apply this ``unmaking'' here to our
case-study of Stable Diffusion.

Stable Diffusion is relatively unique in terms of its ownership and
business structure. Stable Diffusion was developed by CompVIS, a
research group at Ludwig Maximilian University of Munich, LAION, a
non-profit organization based in Germany, and StabilityAI, a
London-based corporation. This collaborative public/private partnership
is significantly different from the typical Silicon Valley model in
which a single corporate entity develops, ships, and maintains a piece
of software or a service. Rather than the established binary of industry
or academia---a startup with business pressures or a lab lacking capital
and compute resources--- the organizational structure here is something
else, a hybrid model (Mostaque 2022).

Collaborating with public institutions has generally resulted in
openness in two senses. First, there has been full transparency
regarding the model's source code, production process, and datasets,
which are publicly available. The model itself is published under a Open
Responsible AI License ``designed to permit free and open access,
re-use, and downstream distribution of derivatives of AI artifacts as
long as the behavioral-use restrictions always apply'' (Ferrandis 2023).
While these behavioral restrictions are interesting, placing the onus on
the user for responsible use, the key point in this context is the open
license, which broadly follows the ideals of the free and open source
software movements (DiBona and Ockman 1999; Ebert 2008) in allowing
users to adapt, modify, and build on software as an expression of their
civil liberties. This differs significantly from competitors like DALL-E
and MidJourney, for example, who offer a closed-source service with
specific terms and conditions which differ significantly between paid
and free accounts.

Secondly, openness can be seen in the ecosystem that surrounds Stable
Diffusion. Along with the official Discord servers, there is a ``Not
Safe for Work'' server, a Reddit group with 170,000 subscribers, a
bewildering number of model versions on Github, each with their own
groups, and associated pages on HuggingFace (2023), the ``AI community
building the future.'' This ecosystem is sprawling and rapidly-evolving,
generating numerous variants of the model, with different features and
add-ons, for different use-cases. If MidJourney is a high-end gallery,
with a carefully curated single model, Stable Diffusion can be a kind of
wild west, a largely unregulated terrain where many actors are using and
abusing the model to suit their own purposes (Vincent 2022). This
openness is amplified by the fact that the model, in various modified
forms, can be downloaded, operated and modified, with some technical
expertise, on consumer hardware. Moreover, even the ``official'' version
of the model maintained by StabilityAI remains receptive to this
permeable ecosystem. Their Discord server actively engages with internet
users and their release notes take pride in listening to users and
fine-tuning the model accordingly (Stability AI 2022). This active
ecosystem is what StabiliyAI's founder calls its ``community vibe or
community structure,'' (Mostaque 2022) and is intentionally cultivated
by the company through various efforts.

For some, these moves represent the ``democratization of AI'' (Shimizu
2022) or as Stability AI (2023) frames it, ``AI by the people and for
the people.'' In terms of image production, this ostensible
democratization seems to be double-edged. On the one hand, it produces a
``rougher'' model, swinging wildly in terms of results, requiring more
hand-holding, and differing significantly from version to version. This
characteristic becomes even clearer when compared against MidJourney and
DALL-E, for instance, which both use a single model, carefully designed
and curated by a single company, which produces ``professional'' results
with minimal prompting.

On the other hand, this democratization defers considerable
responsibility for image production. Stable Diffusion is taken up,
adapted, and employed by a wide variety of communities for
``legitimate'' and ``illegitimate'' uses (Vincent 2022). Fine-tuned
versions of the model -- typically exhibiting the preoccupations of a
heterosexual male gaze with anime, pornography and fantasy heroines --
are hosted on websites such as Civitai. Other community-led initiatives
have aimed to make Stable Diffusion more flexible, through customized
user interfaces (Automatic1111, ComfyAI, InvokeAI) and extensions for
animation, pose control and fine-tuning (Deforum, ControlNet,
DreamBooth).

The deluge of derivative models, user interfaces, and plugins, alongside
an explosion of social media commentary and YouTube tutorials, produces
its own intoxicating and bewildering object---impossible to survey,
fast-moving, and elusive. This open and extensible dynamic allows
StabilityAI to distance itself from the use of its model ``in the
wild,'' in contrast to other AI models which are closely associated with
their parent companies. In this framing, Stable Diffusion becomes an
internet-based tool, which can be used and abused by ``the people,''
rather than a corporate product, where responsibility is clear, quality
must be ensured, and toxicity must be mitigated.

Unmaking the ecosystem provides a fuller portrait of the model itself.
It shows how key model characteristics and tendencies emerge from the
forces that surround it, whether these are business values, revenue
models, or community-engagement programs. The Silicon Valley model is a
very particular kind of organizational and business model (Aoki and
Takizawa 2002). Similarly Big Tech embodies a very particular set of
values, norms, and cultural imperatives (Birch and Bronson 2022). For
boosters, these structures, values, and visions have led to disruptive
innovations and brilliant technologies (Brynjolfsson and McAfee 2014).
For others, they have produced predatory products and amplified
irresponsible practices, leading to a wider societal disenchantment and
``techlash'' (Heaven 2018). Our point here is not to champion or condemn
certain technologies, but instead to highlight how they are powerfully
shaped by the economic, social, and cultural forces that surround them.
Analyzing this environment can provide insights into a model, why it is
the way it is, and in what ways it differs from other models in the same
space.

\hypertarget{section-3}{%
\subsubsection{}\label{section-3}}

\hypertarget{unmaking-the-data}{%
\subsection{\texorpdfstring{Unmaking the Data
}{Unmaking the Data }}\label{unmaking-the-data}}

Training data is fundamental to contemporary AI models---and this makes
the investigation of that data fundamental to any critical analysis.
Training data is the ``ground truth'' of machine learning, the
underlying reality that models aim to attain and are constantly measured
against. This epistemic foundation implies a sense of stability, a layer
of firm evidence drawn from objective observation (Gil-Fournier and
Parikka 2021). Yet as Bowker (2009) asserted: there is no such thing as
raw data; data must be carefully cooked. These are design decisions and
in this sense, ``the designer of a system holds the power to decide what
the truth of the world will be as defined by a training set'' (Crawford
2022).

Unmaking the data pursues a series of fundamental but consequential
questions. Where is this training data sourced from? What kinds of
shaping, curating, or censoring has it undergone? What is included (and
excluded) in this dataset? Who were the creators and curators of this
material, and what kinds of motivations or interests might they have?
Just as Boyd and Crawford (2012) staged a series of critical questions
for big data, we need to ask these same questions of the training data
that underpins AI models.

Such questions, while basic, are not necessarily easy to answer in the
domain of machine learning, where data sourcing, cleaning, and curation
is generally denigrated as low-level work. The dataset, trivial and
assumed, is gestured to in a few lines, while the model architecture and
production chain is carefully laid out across many pages. Certainly data
sets are not always opaque; there is now ample literature on the
composition of large data sets, with discussions on data availability,
data selection, and data biases (Deng et al. 2009; Gao et al. 2020;
Schuhmann et al. 2022). However, in contrast to earlier foundational
papers on image processing and machine learning, where data sets and
model architectures were presented together (e.g. LeCun et al. 1998;
Krizhevsky \& Hinton 2009), more recent literature has split these
elements---with dataset papers generally receiving less citations. For
example, the paper describing LAION-5B by Schuhmann et al. (2022) -- a
critical data set used by Stable Diffusion, and likely involved in
training its main competitors, Dall-E and MidJourney -- has been cited
353 times, while the paper describing Dall-E 2 by Ramesh et al. (2022)
-- released in the same year, and less influential in practice than
either Stable Diffusion or Midjourney -- has been cited 1,695 times
(Google 2023). While breakthroughs in image and language models have
been met by enormous scholarly and media attention, the work of
compiling data sets, no less critical to the performance of generative
AI systems, is comparatively understated and unlauded. This undervaluing
shrouds training data and motivates work on developing more detailed
data provenance to pave the way for more responsible models (Song and
Shmatikov 2019; Werder et al. 2022).

The training data for Stable Diffusion comes from the massive LAION-5B
(Schuhmann et al. 2022), a dataset consisting of 5 billion image-text
pairs. To create the dataset, developers drew on Common Crawl, an
immense repository of web pages scraped from the internet over twelve
years which is now petabytes in size. Developers identified all the
image tags in each webpage along with the ``alt-text'' or description
that typically accompanies it. These images were classified by language
and segmented into different datasets based on common properties
(resolution, probable existence of watermark, etc). Stable Diffusion was
initially trained on a 2-billion item subset of this data, while the
last few trainings have been on LAION Aesthetics 2.5+, a dataset without
watermarks, without low-resolution images, and with a predicted
``aesthetic'' score of 5 or higher (Baio 2022).

There are a number of widely documented issues with this dataset; here
we focus on just three. First, intellectual property. LAION links to
millions of copyrighted images, including those in stock libraries, in
government databases, and in institutional archives. The creators of
LAION always maintained that the dataset was not designed for commercial
use, and distanced themselves from any copyright violations by stressing
that the dataset only contains links to images on the internet, rather
than ``containing'' the images themselves (Beaumont 2022; Schuhmann et
al. 2022). Despite that, StabilityAI and other companies have leveraged
LAION for live products like Stable Diffusion, harvesting all these
image links and using them to train the model. There are clear copyright
issues here (Webster 2023; Strowel 2023), with a number of corporations
and artists now bringing lawsuits against the model and its parent
company. However, as the model is now widely deployed, such regulation
may be a case of ``too little, too late'' -- a recurring dilemma with
fast-moving technology and slow-moving legislative reforms (Moses 2007).

The second issue, closely related, is artistic labor. A cursory search
in LAION for images reveals millions of images that are clearly the work
of artists, photographers, illustrators, and designers. The dataset
contains many images from Deviant Art, an online art community
containing a mixture of amateur homages and more ``professional''
productions. The dataset also houses many images sourced directly from
Fine Art America (2023), a marketplace where ``hundreds of thousands of
artists, photographers, and national brands sell their artwork.''
Indeed, by sorting from highest to lowest in the Aesthetic column of the
database, we can see that oil paintings consistently receive some of the
highest ``aesthetic'' scores. This means they are certainly included in
the Aesthetics subset used to train Stable Diffusion. In fact, a
frequent prompting tip for SD is to include the names of particular
artists in the prompt (Mitchell 2022). Names like H.R. Giger and Ismail
Inceoglu offer a kind of shortcut to achieving a distinct visual style
for a particular genre like landscapes or horror. By harvesting an
artist's work en-masse without their consent, models like SD are able to
``identify'' its trademark features and then use it to automatically
generate new images in this style. Websites like haveibeentrained (2023)
allow artists to search for their work within the LAION dataset and then
``opt out'' of future models. However, this mechanism only applies to
future models and assumes that companies are using the ``official''
application programming interface (API) to access the model.

The third issue is toxic content, broadly conceived. Examining LAION
reveals that it contains links to hateful and racist material, to
depictions of cruelty, and to millions of items of pornography. These
findings have been echoed in academic research (Qu et al. 2023) and in
investigative journalism, which identified terrorist executions and
non-consensual porn in the dataset (Xiang and Maiberg 2022;). There is a
legislative aspect here, where some of this material, depending on the
jurisdiction it was accessed in, would be strictly illegal. But there is
also a normative aspect, which is fuzzier or more subjective. Some of
this material would be considered unethical, immoral, or objectionable,
depending on the viewer. But the threshold (or appetite) for such
material varies greatly depending on the society, culture, religion, and
broader context. There is also a double-edged quality here to imagery
depicting the human body, allowing for very different uses. For
instance, one generative AI company discovered that users were using the
model to generate violent and hypersexualized imagery for their personal
needs, and banned a number of keywords to prevent this. However, this
move also prevented ``legitimate'' prompts such as medical illustrations
that aimed to depict female reproduction (Heikkilä 2023). Other users
have criticized developers for ``censoring'' content and creating
prudish systems (Bratton 2023). These tensions underscore the darker and
more controversial content contained within datasets and the
difficulties (or impossibilities) in arriving at a societal consensus
regarding the acceptable and unacceptable.

There are certainly other important issues which could be pursued around
this image material. Stable Diffusion developers, for instance,
acknowledge the potential for algorithmic bias, given the predominance
of Western, English-speaking descriptions and the under-representation
of data on other languages, cultures, and perspectives (Rombach and
Esser 2022). However, the three issues flagged here serve to showcase
the productive potential of the line of inquiry.

\hypertarget{unmaking-the-output}{%
\subsection{Unmaking the Output}\label{unmaking-the-output}}

The third method focuses on the output of the model, interrogating it
through a process of prompting, reflection, and iteration. In essence,
this method is a slower and more reflexive version of the average user
experience. As with other digital methods (Light et al. 2018; van Geenen
2020), this method adopts the perspective of the end-user to gain
insights about a particular technical system or online environment.
However, rather than gradually arriving at a desirable or
``professional'' image, this prompting and iteration process aims to
reveal the logics of image production, including their inconsistencies,
edge cases, and break-down points. In other words, this method aims for
seamfulness (Chalmers and MacColl 2003; Wenneling 2007) rather than
seamlessness, disrupting the smooth experience and taken-for-granted
output that tends to dominate our experience of contemporary
technologies.

Like the other methods, then, this method is not a step-by-step recipe
but a mode of inquiry that asks certain questions, carries out certain
activities, and maintains a certain sensibility when reflecting on the
results. This AI image reading might be carried out with any number of
models or topics (Salvaggio 2022). However, in our case, we use a mix of
Stable Diffusion 1.5 using DiffusionBee (an open source application run
on a local machine) and SDXL 0.9, using DreamBooth (a web-based
commercial offering by StabilityAI). We use two deliberately simplified
prompts to show ``default'' features and operations of the model, using
common historically gendered occupations of ``lawyer'' and ``nurse.''

We select this example not for its novelty but precisely because it is
well-known. Indeed, the reproduction of gender norms has been a
canonical case of bias repeatedly tested over the years with the latest
AI models. Bolukbasi et al. (2017), for example, in their paper titled
``man is to computer programmer as woman is to homemaker,'' observed
that word embeddings in AI models exhibited gendered stereotypes to a
disturbing extent. This study has been followed by others which
investigate gendered inaccuracies in models (Buolamwini and Gebru 2018),
confirm systemic gendered bias in some systems (Rudinger et al. 2018),
and explore how these forms of bias intersect with other such as
religion and disability (Magee et al. 2021).

The image grids generated by the prompt ``portrait of a lawyer'' in
Figures 1 and 2 below immediately highlight a particular kind of visual
representation associated with this prestigious and profitable position:
white males. These individuals are generally middle-aged to more senior
in appearance. They all wear formal legal gowns or business suits with
ties. Even the figure in the lower right corner of Figure 1, while
having some traditionally ``feminine'' qualities (longer hair, narrower
shoulders), arguably presents as male. And while the more recent SDXL
model includes several figures of color, these remain resolutely male.
These results resonate with recent findings on popular image models such
as Stable Diffusion 1.5, Stable Diffusion 2.1, and Dall-E 2. After
analyzing over 96,000 images, for instance, Luccioni et al. (2023) found
that all three models significantly over-represented the portion of
their latent space associated with whiteness and masculinity.

\begin{figure*}
    \centering
    \includegraphics[width=3in,height=3in]{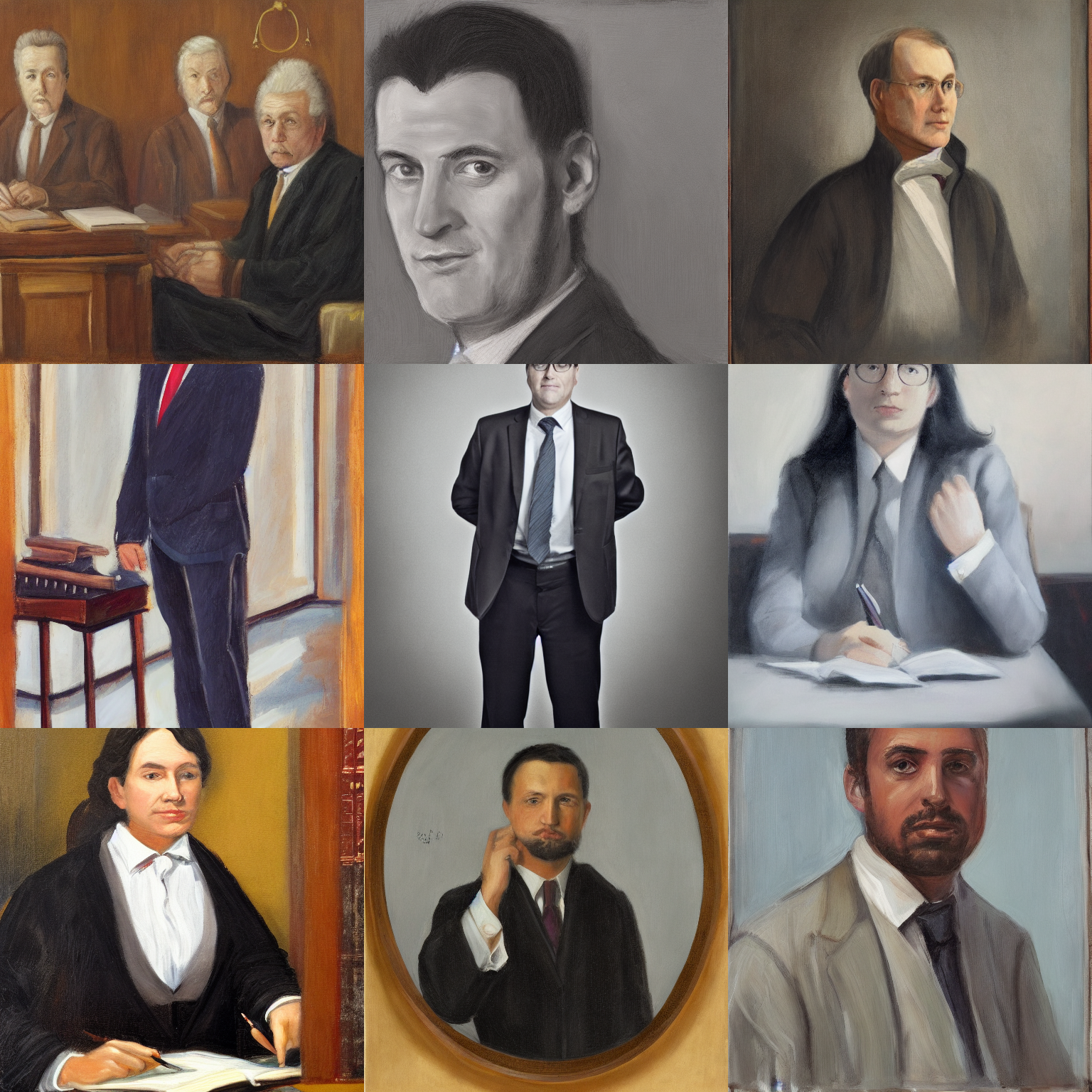}
    \label{fig:fig1}
    \caption{``Lawyer'' StableDiffusion 1.5}
\end{figure*}

\begin{figure*}
    \centering
    \includegraphics[width=3in,height=3in]{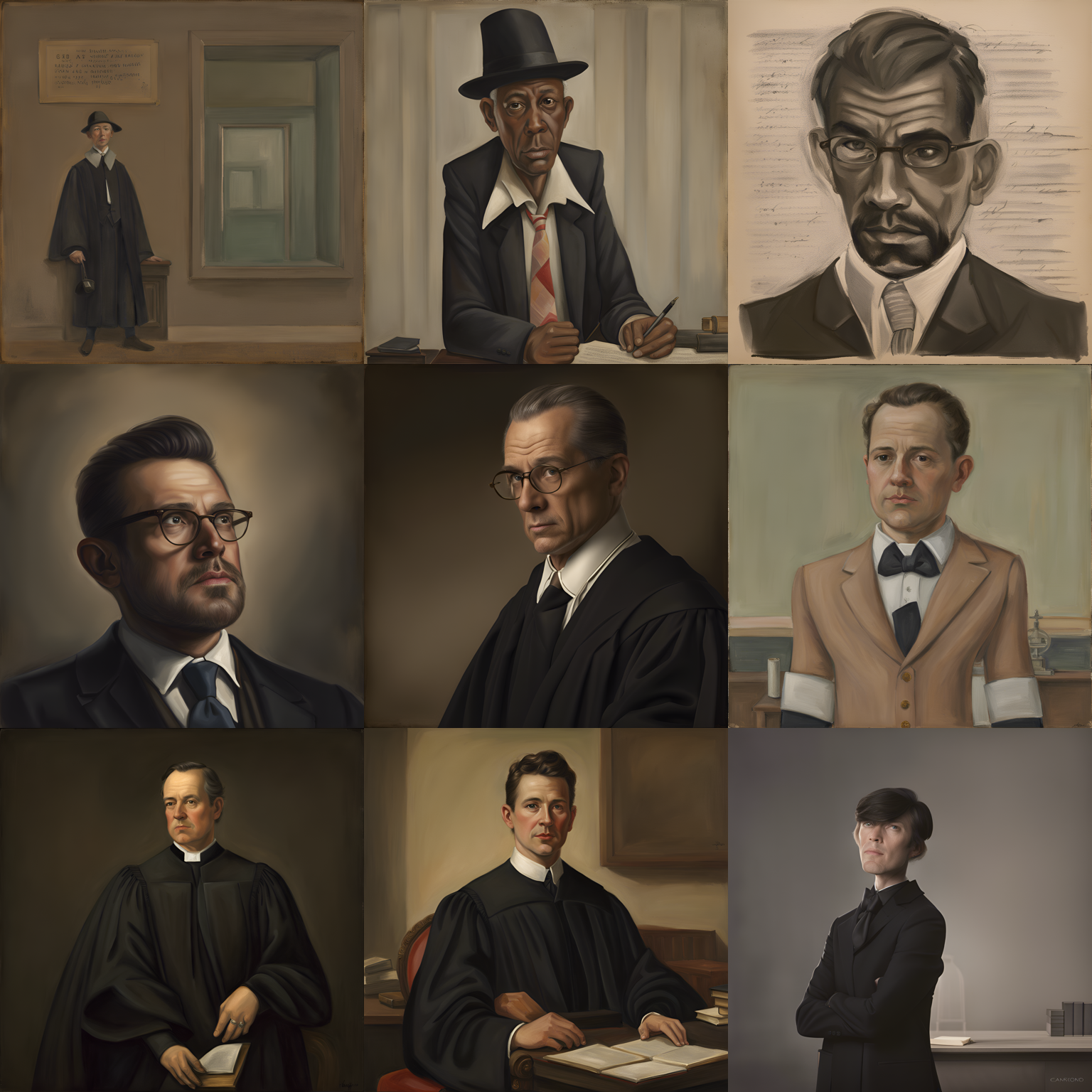}
    \label{fig:fig1}
    \caption{``Lawyer'' SDSXL 0.9}
\end{figure*}

\begin{figure*}
    \centering
    \includegraphics[width=3in,height=3in]{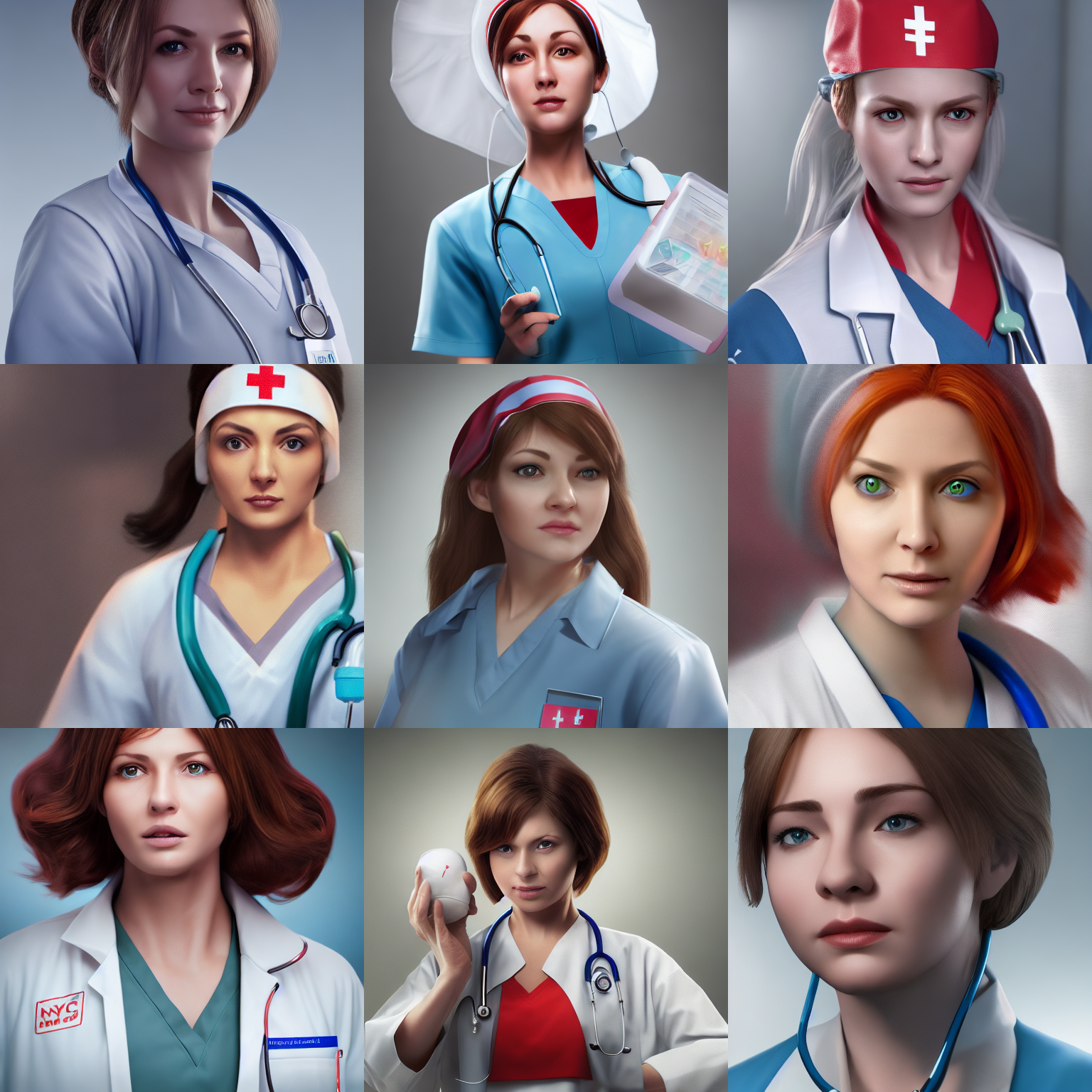}
    \label{fig:fig1}
    \caption{``Nurse'' StableDiffusion 1.5}
\end{figure*}

\begin{figure*}
    \centering
    \includegraphics[width=3in,height=3in]{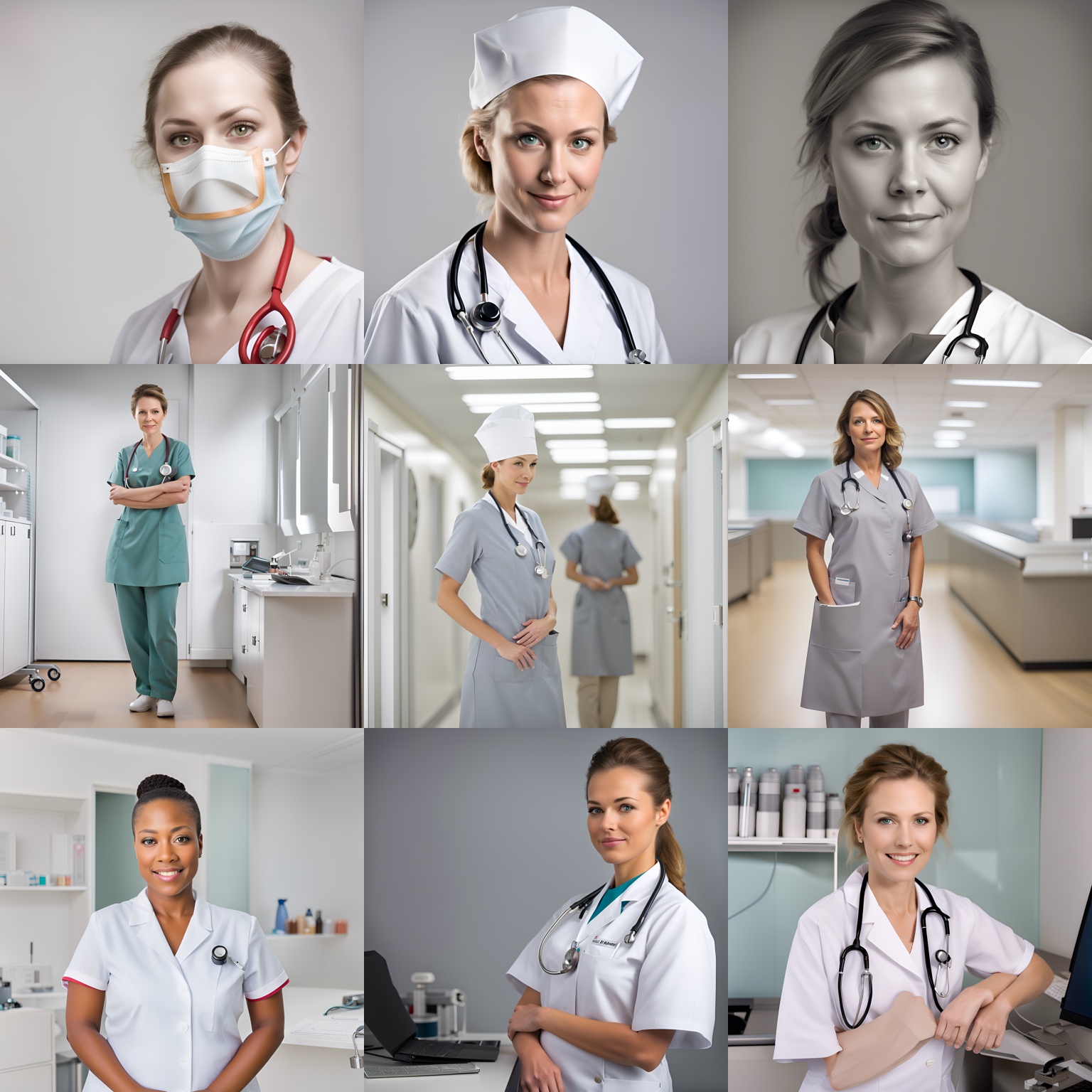}
    \label{fig:fig1}
    \caption{``Nurse'' SDXL 0.9}
\end{figure*}

The second set of images is similarly based on a simple prompt: ``nurse,
hyper realistic, 4k'' for Figure 3 (showing images generated by Stable
Diffusion 1.5), and a slightly more expansive form, ``A photo of a
nurse, professional photography, nikon, highly detailed'' for Figure 4
(images generated by SDXL 0.9). The first immediately apparent point
concerns the nurse as a figure. Just as with the lawyer example,
``nurse'' is stereotyped in a heavily gendered way, with all nine images
being female. In addition, most of these figures present as white
European, with the exception of one or two standouts which have some
Latina, Middle Eastern, African or African-American features. Perhaps
most striking is that all of the faces are youthful and classically
beautiful with smooth skin, defined features, and symmetrical
structures. From an aesthetic perspective, we can also ascertain, for
the first set, that the ``hyperrealistic'' and ``4k'' phrases produce a
hybrid result that sits somewhere between highly stylized photograph,
digital art, and 3d render. The second set showcases, with some
additional prompt hints, the photorealism possible in Stable Diffusion's
updated model. Again the figures are youthful, and caught, as the prompt
requests, in a set of poses specific to the genre of portraiture.

Such results resonate with other research on AI image models. In a
series of experiments around sexualization and generative models, Wolfe
et al. (2023) found that images of female professionals (scientists,
doctors, executives) are likely to be associated with sexual
descriptions relative to images of male professionals. More generally,
Fraser et al. (2023) carried out a number of underspecified prompts such
as ``threatening face'' vs ``friendly face'' and found that, in many
cases, generated images contained stereotypes and demographic biases.
And moving beyond the binaries of men and women, Ungless et al. (2023)
found that certain non-cisgender identities were consistently
(mis)represented as less human, more stereotyped and more sexualised.

Such experiments are not necessarily about definitely ``proving'' that
image models are biased or stereotyped, but instead about opening up a
longer critical discussion and programme. These outputs and the research
cited above point to a series of thorny issues at the core of generative
image models. While such issues certainly include gendered stereotypes,
they also gesture to a wider array of issues around power, difference,
and representation. Whose image are we most often generating and whose
images are ``missing'' or under-represented? Which kinds of images are
associated with aspirational roles and which images are linked to more
pathological prompts? Digging into these questions requires digging into
the tool itself through longer sessions of prompting, reflection, and
iteration--- and this is the kind of sustained engagement that we see as
necessary for a critical understanding of contemporary AI image models.

\hypertarget{conclusion-new-machines-new-methods}{%
\subsection{Conclusion: New Machines, New
Methods}\label{conclusion-new-machines-new-methods}}

AI image models are powerful in their ability to synthesize a wide array
of image material based on natural language prompts. This ability has
seen them become rapidly adopted across a variety of industries
(Kulkarni et al. 2023; Park et al. 2022). However, these abilities are
also attended by major issues, including the perpetuation of stereotypes
and biases (Birnhane 2021; Luccioni 2023). These issues have downstream
impacts, shaping representation and aesthetic production in significant
ways and posing important questions to society.

However, the public's understanding of these models, from their
development to their training data and their operational logics, is
often limited (Emmert-Streib et al. 2020). It is urgently necessary,
therefore, to develop frameworks, methods, and concepts to better grasp
generative image models and their work in high-stakes areas. The point
here is to not just ``correct'' fallacies, but rather to provide a more
nuanced and multifaceted portrait of these models in order to show their
powerful (and problematic) work in the world. The aim of a
methodological toolkit is not to merely obtain more information, but to
act as a foundation for a broader political project of awareness,
critique, regulation, and even resistance.

To support this work, we provided an accessible framework for critically
analyzing AI image models. Rather than a single method from a single
discipline, we intentionally bring together three methodological
approaches drawn from disparate disciplines. \emph{Unmaking the
ecosystem} analyzed the values, structures, and incentives surrounding
the model's production. \emph{Unmaking the data} analyzed the images and
text the model draws upon, with their attendant particularities and
biases. And \emph{Unmaking the output} analyzed the model's generative
results, revealing its logics through prompting, reflection, and
iteration. Together, these modes of inquiry construct a sociotechnical
portrait of the image model that account for its influences,
architectures, and operational logics. Such socially- and
politically-attuned analyses of generative AI image models pave the way
for a better understanding of their work in the world.

\hypertarget{section-4}{%
\subsection{}\label{section-4}}

\hypertarget{section-5}{%
\subsection{\texorpdfstring{\hfill\break
}{ }}\label{section-5}}

\hypertarget{references}{%
\subsection{References}\label{references}}

Abduljawad, Mohamed, and Abdullah Alsalmani. 2022. ``Towards Creating
Exotic Remote Sensing Datasets Using Image Generating AI.'' In
\emph{2022 International Conference on Electrical and Computing
Technologies and Applications (ICECTA)}, 84--88.
\url{https://doi.org/10.1109/ICECTA57148.2022.9990245}.

Aoki, Masahiko, and Hirokazu Takizawa. 2002. ``Information, Incentives,
and Option Value: The Silicon Valley Model.'' \emph{Journal of
Comparative Economics} 30 (4): 759--86.
\url{https://doi.org/10.1006/jcec.2002.1804}.

Azar, Mitra, Geoff Cox, and Leonardo Impett. 2021. ``Introduction: Ways
of Machine Seeing.'' \emph{AI \& SOCIETY}, 1--12.

Baio, Andy. 2022. ``Exploring 12 Million of the 2.3 Billion Images Used
to Train Stable Diffusion's Image Generator.'' \emph{Waxy.Org} (blog).
August 30, 2022.
\url{https://waxy.org/2022/08/exploring-12-million-of-the-images-used-to-train-stable-diffusions-image-generator/}.

Beaumont, Romain. 2022. ``Laion-5b: A New Era of Open Large-Scale
Multi-Modal Datasets.'' March 31, 2022.
\url{https://laion.ai/blog/laion-5b/}.

Benjamin Bratton {[}@bratton{]}. 2023. ``For All the By-the-Numbers
Articles on Potential Harms of Representing Bad Ideas Words and Things
with Generative LLM, I Am Surprised by Dearth of Discussion of
Lobotomized, Prudish, Death-Phobic, Sex-Phobic Public Models Now
Foundational. Who's Writing Smart Things on This?'' Tweet.
\emph{Twitter}.
\url{https://twitter.com/bratton/status/1618391158388060161}.

Benjamin, Walter. 1935. ``The Work of Art in the Age of Mechanical
Reproduction, 1936.''

Biderman, Stella, Kieran Bicheno, and Leo Gao. 2022. ``Datasheet for the
Pile.'' \emph{ArXiv Preprint ArXiv:2201.07311}.

Birch, Kean, and Kelly Bronson. 2022. ``Big Tech.'' \emph{Science as
Culture} 31 (1): 1--14.
\url{https://doi.org/10.1080/09505431.2022.2036118}.

Birhane, Abeba, Vinay Uday Prabhu, and Emmanuel Kahembwe. 2021.
``Multimodal Datasets: Misogyny, Pornography, and Malignant
Stereotypes.'' arXiv.
\url{https://doi.org/10.48550/arXiv.2110.01963}.

boyd, danah, and Kate Crawford. 2012. ``Critical Questions for Big Data:
Provocations for a Cultural, Technological, and Scholarly Phenomenon.''
\emph{Information, Communication \& Society} 15 (5): 662--79.

Brynjolfsson, Erik, and Andrew McAfee. 2014. \emph{The Second Machine
Age: Work, Progress, and Prosperity in a Time of Brilliant
Technologies}. New York: WW Norton \& Company.

Chalmers, Matthew, and Ian MacColl. 2003. ``Seamful and Seamless Design
in Ubiquitous Computing.'' In \emph{Workshop at the Crossroads: The
Interaction of HCI and Systems Issues in UbiComp}. Vol. 8.

Clarke, Laurie. 2022. ``When AI Can Make Art -- What Does It Mean for
Creativity?'' \emph{The Observer}, November 12, 2022.
\url{https://www.theguardian.com/technology/2022/nov/12/when-ai-can-make-art-what-does-it-mean-for-creativity-dall-e-midjourney}.

Crary, Jonathan. 1992. \emph{Techniques of the observer: On vision and
modernity in the nineteenth century.} Cambridge, MA: MIT press.

Deng, Jia, Wei Dong, Richard Socher, Li-Jia Li, Kai Li, and Li Fei-Fei.
2009. ``Imagenet: A Large-Scale Hierarchical Image Database.'' In
\emph{2009 IEEE Conference on Computer Vision and Pattern Recognition},
248--55. IEEE.

Derrida, Jacques. 1996. \emph{Archive Fever: A Freudian Impression}.
Chicago: University of Chicago Press.

DiBona, Chris, and Sam Ockman. 1999. \emph{Open Sources: Voices from the
Open Source Revolution}. Sebastopol, CA: O'Reilly Media.

Ebert, Christof. 2008. ``Open Source Software in Industry.'' \emph{IEEE
Software} 25 (3): 52--53.
\url{https://doi.org/10.1109/MS.2008.67}.

Emmert-Streib, Frank, Olli Yli-Harja, and Matthias Dehmer. 2020.
``Artificial Intelligence: A Clarification of Misconceptions, Myths and
Desired Status.'' \emph{Frontiers in Artificial Intelligence} 3.
\url{https://www.frontiersin.org/articles/10.3389/frai.2020.524339}.

Facchini, Alessandro, and Alberto Termine. 2022. ``Towards a Taxonomy
for the Opacity of AI Systems.'' In \emph{Philosophy and Theory of
Artificial Intelligence 2021}, edited by Vincent C. Müller, 73--89.
Studies in Applied Philosophy, Epistemology and Rational Ethics. Cham:
Springer International Publishing.
\url{https://doi.org/10.1007/978-3-031-09153-7_7}.

Fraser, Kathleen C., Svetlana Kiritchenko, and Isar Nejadgholi. 2023.
``A Friendly Face: Do Text-to-Image Systems Rely on Stereotypes When the
Input Is Under-Specified?'' arXiv.
\url{https://doi.org/10.48550/arXiv.2302.07159}.

Gal, Rinon, Yuval Alaluf, Yuval Atzmon, Or Patashnik, Amit H. Bermano,
Gal Chechik, and Daniel Cohen-Or. 2022. ``An Image Is Worth One Word:
Personalizing Text-to-Image Generation Using Textual Inversion.'' arXiv.
\url{https://doi.org/10.48550/arXiv.2208.01618}.

Gao, Leo, Stella Biderman, Sid Black, Laurence Golding, Travis Hoppe,
Charles Foster, Jason Phang, Horace He, Anish Thite, and Noa Nabeshima.
2020. ``The Pile: An 800gb Dataset of Diverse Text for Language
Modeling.'' \emph{ArXiv Preprint ArXiv:2101.00027}.

Geenen, Daniela van. 2020. ``Critical Affordance Analysis for Digital
Methods: The Case of Gephi,'' 1--21.
\url{https://doi.org/10.25969/mediarep/14855}.

Ghaffari, Kimia, Mohammad Lagzian, Mostafa Kazemi, and Gholamreza
Malekzadeh. 2019. ``A Socio-Technical Analysis of Internet of Things
Development: An Interplay of Technologies, Tasks, Structures and
Actors.'' \emph{Foresight} 21 (6): 640--53.
\url{https://doi.org/10.1108/FS-05-2019-0037}.

Gil-Fournier, Abelardo, and Jussi Parikka. 2021. ``Ground Truth to Fake
Geographies: Machine Vision and Learning in Visual Practices.'' \emph{AI
\& SOCIETY} 36 (4): 1253--62.
\url{https://doi.org/10.1007/s00146-020-01062-3}.

Google. 2023. ``Google Scholar.'' 2023.
\url{https://scholar.google.com/schhp?hl=en\&as_sdt=0,5}.

Heaven, Douglas. 2018. ``Techlash.'' \emph{New Scientist} 237 (3164):
28--31.
\url{https://doi.org/10.1016/S0262-4079(18)30259-8}.

Heikkilä, Melissa. 2023. ``AI Image Generator Midjourney Blocks Porn by
Banning Words about the Human Reproductive System.'' MIT Technology
Review. February 24, 2023.
\url{https://www.technologyreview.com/2023/02/24/1069093/ai-image-generator-midjourney-blocks-porn-by-banning-words-about-the-human-reproductive-system/}.

Hill, Kashmir. 2023. ``This Tool Could Protect Artists From
A.I.-Generated Art That Steals Their Style.'' \emph{The New York Times},
February 13, 2023.
\url{https://www.nytimes.com/2023/02/13/technology/ai-art-generator-lensa-stable-diffusion.html}.

Kang, Edward B. 2023. ``Ground Truth Tracings (GTT): On the Epistemic
Limits of Machine Learning.'' \emph{Big Data \& Society} 10 (1):
20539517221146120.
\url{https://doi.org/10.1177/20539517221146122}.

Krizhevsky, Alex, and Geoffrey Hinton. 2009. ``Learning Multiple Layers
of Features from Tiny Images.''

Kulkarni, Chinmay, Stefania Druga, Minsuk Chang, Alex Fiannaca, Carrie
Cai, and Michael Terry. 2023. ``A Word Is Worth a Thousand Pictures:
Prompts as AI Design Material.'' ArXiv.Org. March 22, 2023.
\url{https://arxiv.org/abs/2303.12647v1}.

LeCun, Yann, Bernhard Boser, John S Denker, Donnie Henderson, Richard E
Howard, Wayne Hubbard, and Lawrence D Jackel. 1989. ``Backpropagation
Applied to Handwritten Zip Code Recognition.'' \emph{Neural Computation}
1 (4): 541--51.

LeCun, Yann, Léon Bottou, Yoshua Bengio, and Patrick Haffner. 1998.
``Gradient-Based Learning Applied to Document Recognition.''
\emph{Proceedings of the IEEE} 86 (11): 2278--2324.

Light, Ben, Jean Burgess, and Stefanie Duguay. 2018. ``The Walkthrough
Method: An Approach to the Study of Apps.'' \emph{New Media \& Society}
20 (3): 881--900.

Luccioni, Alexandra Sasha, Christopher Akiki, Margaret Mitchell, and
Yacine Jernite. 2023. ``Stable Bias: Analyzing Societal Representations
in Diffusion Models.'' ArXiv.Org. March 20, 2023.
\url{https://arxiv.org/abs/2303.11408v1}.

Luitse, Dieuwertje, and Wiebke Denkena. 2021. ``The Great Transformer:
Examining the Role of Large Language Models in the Political Economy of
AI.'' \emph{Big Data \& Society} 8 (2): 20539517211047736.

Mackenzie, Adrian, and Anna Munster. 2019. ``Platform Seeing: Image
Ensembles and Their Invisualities.'' \emph{Theory, Culture \& Society}
36 (5): 3--22.

Manovich, Lev. 2017a. ``Aesthetics,`Formalism', and Media Studies.''
\emph{Keywords in Media Studies}, 9--12.

---------. 2017b. ``Automating Aesthetics: Artificial Intelligence and
Image Culture.'' \emph{Flash Art International} 316: 1--10.

Mitchell, Nick. 2023. ``GitHub - Gnickm/Stable-Diffusion-Artists:
Curated List of Artists for Stable Diffusion Prompts.'' Accessed June 9,
2023.
\url{https://github.com/gnickm/stable-diffusion-artists}.

Moses, Lyria Bennett. 2007. ``Recurring Dilemmas: The Law's Race to Keep
up with Technological Change.'' \emph{U. Ill. JL Tech. \& Pol'y}, 239.

Mostaque, Emad. \emph{The Man behind Stable Diffusion}.
\url{https://www.youtube.com/watch?v=YQ2QtKcK2dA}.

Myer, David. 2022. ``Stability AI Can't Please Everyone with Stable
Diffusion 2.0.'' Fortune. December 1, 2022.
\url{https://fortune.com/2022/11/30/stable-diffusion-2-stability-ai-artists-nsfw-celebrities-copyright/}.

Nieborg, David B, and Anne Helmond. 2019. ``The Political Economy of
Facebook's Platformization in the Mobile Ecosystem: Facebook Messenger
as a Platform Instance.'' \emph{Media, Culture \& Society} 41 (2):
196--218.

Offert, Fabian, and Thao Phan. 2022. ``A Sign That Spells: DALL-E 2,
Invisual Images and The Racial Politics of Feature Space.'' \emph{ArXiv
Preprint ArXiv:2211.06323}.

Paglen, Trevor. 2014. ``Operational Images - Journal \#59 November 2014
- e-Flux.'' November 2014.
\url{http://www.e-flux.com/journal/59/61130/operational-images/}.

Parisi, Luciana. 2021. ``Negative Optics in Vision Machines.'' \emph{AI
\& SOCIETY} 36: 1281--93.

Park, Ji Eun, Philipp Vollmuth, Namkug Kim, and Ho Sung Kim. 2022.
``Research Highlight: Use of Generative Images Created with Artificial
Intelligence for Brain Tumor Imaging.'' \emph{Korean Journal of
Radiology} 23 (5): 500.

Qu, Yiting, Xinyue Shen, Xinlei He, Michael Backes, Savvas Zannettou,
and Yang Zhang. 2023. ``Unsafe Diffusion: On the Generation of Unsafe
Images and Hateful Memes From Text-To-Image Models.'' ArXiv.Org. May 23,
2023.
\url{https://arxiv.org/abs/2305.13873v1}.

Ramesh, Aditya, Prafulla Dhariwal, Alex Nichol, Casey Chu, and Mark
Chen. 2022. ``Hierarchical Text-Conditional Image Generation with Clip
Latents.'' \emph{ArXiv Preprint ArXiv:2204.06125}.

Rombach, Robin, and Patrick Esser. 2022. ``CompVis/Stable-Diffusion-v1-4
· Hugging Face.'' 2022.
\url{https://huggingface.co/CompVis/stable-diffusion-v1-4}.

Salvaggio, Eryk. 2022. ``How to Read an AI Image.'' \emph{Cybernetic
Forests}. October 2, 2022.
\url{https://cyberneticforests.substack.com/p/how-to-read-an-ai-image}.

Schuhmann, Christoph, Romain Beaumont, Richard Vencu, Cade Gordon, Ross
Wightman, Mehdi Cherti, Theo Coombes, Aarush Katta, Clayton Mullis, and
Mitchell Wortsman. 2022. ``Laion-5b: An Open Large-Scale Dataset for
Training next Generation Image-Text Models.'' \emph{ArXiv Preprint
ArXiv:2210.08402}.

Seger, Elizabeth, Aviv Ovadya, Ben Garfinkel, Divya Siddarth, and Allan
Dafoe. 2023. ``Democratising AI: Multiple Meanings, Goals, and
Methods.'' arXiv.
\url{https://doi.org/10.48550/arXiv.2303.12642}.

Sekula, Allan. 2003. ``Reading an Archive: Photography between Labour
and Capital.'' \emph{The Photography Reader}, 443--52.

Shimizu, Ryo. 2022. ``Beyond Midjourney? Reason Why Free Drawing AI
`\#StableDiffusion' Can Assert That `AI Has Been Democratized.'\,''
Business Insider. August 26, 2022.
\url{https://www.businessinsider.jp/post-258369}.

Song, Congzheng, and Vitaly Shmatikov. 2019. ``Auditing Data Provenance
in Text-Generation Models.'' In \emph{Proceedings of the 25th ACM SIGKDD
International Conference on Knowledge Discovery \& Data Mining},
196--206.

Stability AI. 2022. ``Stable Diffusion v2.1 and DreamStudio Updates.''
Stability AI. December 7, 2022.
\url{https://stability.ai/blog/stablediffusion2-1-release7-dec-2022}.

Strowel, Alain. 2023. ``ChatGPT and Generative AI Tools: Theft of
Intellectual Labor?'' \emph{IIC - International Review of Intellectual
Property and Competition Law} 54 (4): 491--94.
\url{https://doi.org/10.1007/s40319-023-01321-y}.

Ungless, Eddie L., Björn Ross, and Anne Lauscher. 2023. ``Stereotypes
and Smut: The (Mis)Representation of Non-Cisgender Identities by
Text-to-Image Models.'' ArXiv.Org. May 26, 2023.
\url{https://arxiv.org/abs/2305.17072v1}.

Vertesi, Janet, David Ribes, Laura Forlano, Yanni Loukissas, and Marisa
Leavitt Cohn. 2016. ``Engaging, Designing, and Making Digital Systems.''
In \emph{The Handbook of Science and Technology Studies}, 169--94.
Cambridge, MA: MIT Press.

Vincent, James. 2022. ``Anyone Can Use This AI Art Generator --- That's
the Risk.'' The Verge. September 15, 2022.
\url{https://www.theverge.com/2022/9/15/23340673/ai-image-generation-stable-diffusion-explained-ethics-copyright-data}.

Wasielewski, Amanda. 2023. \emph{Computational Formalism: Art History
and Machine Learning}. Cambridge, MA: MIT Press.

Webster, Ryan. 2023. ``A Reproducible Extraction of Training Images from
Diffusion Models.'' arXiv.
\url{https://doi.org/10.48550/arXiv.2305.08694}.

Wenneling, Oskar. 2007. ``Seamful Design--the Other Way Around.'' In
\emph{Proceedings of the Scandinavian Student Interaction Design
Research Conference}, 14--16.

Werder, Karl, Balasubramaniam Ramesh, and Rongen (Sophia) Zhang. 2022.
``Establishing Data Provenance for Responsible Artificial Intelligence
Systems.'' \emph{ACM Transactions on Management Information Systems} 13
(2): 22:1-22:23.
\url{https://doi.org/10.1145/3503488}.

Wolfe, Robert, Yiwei Yang, Bill Howe, and Aylin Caliskan. 2022.
``Contrastive Language-Vision AI Models Pretrained on Web-Scraped
Multimodal Data Exhibit Sexual Objectification Bias.'' ArXiv.Org.
December 21, 2022.
\url{https://doi.org/10.1145/3593013.3594072}.

Woodhouse, Edward, and Jason W. Patton. 2004. ``Introduction: Design by
Society: Science and Technology Studies and the Social Shaping of
Design.'' \emph{Design Issues} 20 (3): 1--12.

Xiang, Chloe, and Emanuel Maiberg. 2022. ``ISIS Executions and
Non-Consensual Porn Are Powering AI Art.'' \emph{Vice} (blog). September
21, 2022.
\url{https://www.vice.com/en/article/93ad75/isis-executions-and-non-consensual-porn-are-powering-ai-art}.

\end{multicols*}

\end{document}